\documentclass[iop]{emulateapj}
\usepackage{amsmath} 
\usepackage{amssymb}
\newcommand{\uv}{\mbox{$u$-$v$}}

\newcommand{\muas}{\mbox{$\mu$as}}

\newcommand{\kms}{\mbox{km s$^{-1}$}}
\newcommand{\Jb}{\mbox{Jy beam$^{-1}$}}
\newcommand{\muJb}{\mbox{$\mu$Jy beam$^{-1}$}}
\newcommand{\RA}[4]{\mbox{${#1}^{\rm h} \; {#2}^{\rm m} \; {#3}\fs{#4} $}}
\newcommand{\dec}[4]{\mbox{${#1}\arcdeg \; {#2}\arcmin \; {#3}\farcs{#4} $}}
\newcommand{\pk}{_{\rm op}}   

\shortauthors{Bietenholz et al.}
\shorttitle{VLBI Observations of SN 2011\lowercase{dh}}

\begin{document}
      
\title{VLBI Observations of the Nearby Type IIb Supernova 2011\lowercase{dh}}

\author{M. F. Bietenholz\altaffilmark{1,2},
A. Brunthaler\altaffilmark{3,4}, A. M. Soderberg\altaffilmark{5},
M. Krauss\altaffilmark{4}, B. Zauderer\altaffilmark{5},
N. Bartel\altaffilmark{2},
L. Chomiuk\altaffilmark{4,5,6}, and
M. P. Rupen\altaffilmark{4}}

\altaffiltext{1}{Hartebeesthoek Radio Observatory, PO Box 443, Krugersdorp, 
1740, South Africa} 
\altaffiltext{2}{Department of Physics and Astronomy, York University, Toronto,
M3J~1P3, Ontario, Canada}
\altaffiltext{3}{Max-Planck-Institut f\"ur Radioastronomie, Auf dem H\"ugel 69, 
53121 Bonn, Germany}
\altaffiltext{4}{National Radio Astronomy Observatory, Socorro, NM 87801, USA}
\altaffiltext{5}{Harvard-Smithsonian Center for Astrophysics, 60
Garden Street, Cambridge, MA 02138, USA}
\altaffiltext{6}{Jansky Fellow of the National Radio Astronomy Observatory}

\slugcomment{Accepted to the Astrophysical Journal}

\begin{abstract}
We report on phase-referenced VLBI radio observations of the Type IIb
supernova 2011dh, at times $t = 83$~days and 179~days after the
explosion and at frequencies, respectively, of 22.2 and 8.4~GHz.  We
detected SN~2011dh at both epochs.  At the first epoch only an upper
limit on SN~2011dh's angular size was obtained, but at the second
epoch, we determine the angular radius SN~2011dh's radio emission to
be $0.25 \pm 0.08$~mas by fitting a spherical shell model directly to
the visibility measurements.  At a distance of 8.4~Mpc this angular
radius corresponds to a time-averaged (since $t = 0$) expansion
velocity of the forward shock of $21000 \pm 7000$~\kms.  Our measured
values of the radius of the emission region are in excellent agreement
with those derived from fitting synchrotron self-absorbed models to
the radio spectral energy distribution, providing strong confirmation
for the latter method of estimating the radius.  We find that
SN~2011dh's radius evolves in a power-law fashion, with $R \propto
t^{0.92 \pm 0.10}$.
\end{abstract}

\keywords{supernovae: individual (SN2011dh) --- radio continuum:
general}

\section{Introduction}
\label{sintro}

Supernova \objectname{SN 2011dh} was discovered on 2011 May 31 by the
amateur astronomer Am\'ad\'ee Riou in the nearby galaxy M51
\citep{Griga+2011}, which is at a distance of $8.4 \pm 0.6$~Mpc
\citep{FeldmeierCJ1997,Vinko+2012}.
The supernova was soon confirmed using pre- and post-discovery imaging
\citep{Griga+2011}.  In particular the optical transient was also
detected by the Palomar Transient Factory \citep{Law+2009} shortly
after the initial discovery. The explosion date is tightly constrained
to be between 2011 May 31.275 and 31.893 UT \citep{Arcavi+2011b}. We
will adopt the (rounded) midpoint of this interval of
$t_0 =$ May 31.6 UT.  The supernova was coincident with the eastern
spiral arm of M51, and had an estimated apparent magnitude of $\sim
14$ mag (unfiltered) implying an absolute magnitude of roughly $\sim
-16$ mag.  We note that M51 is also host to an earlier supernova,
SN~1994I, which however was of Type Ib/c.

Initially, SN~2011dh was spectroscopically classified as Type IIP
\citep{SilvermanFC2011},
but further spectroscopy, which showed helium absorption features,
caused a re-classification as a Type IIb \citep{Arcavi+2011b,
  Marion+2011}.  A maximum expansion velocity of $\sim$17000~\kms\ was
estimated from the blue edge of the H$\alpha$ line
\citep{SilvermanFC2011, Arcavi+2011b}.

Radio emission was detected on June 4, only a few days after the
explosion, at centimeter wavelengths \citep{Horesh+2011} with the
National Radio Astronomy Observatory\footnote{The National Radio
Astronomy Observatory is a facility of the National Science Foundation
operated under cooperative agreement by Associated Universities, Inc.}
(NRAO) Expanded Very Large Array \citep[EVLA;][]{Perley+2011},
at millimeter wavelengths \citep{HoreshZC2011} using the Combined
Array for Research in Millimeter-wave Astronomy (CARMA), and
at sub-mm wavelengths \citep{Soderberg+SN2011dh-I} using the
Submillimeter Array (SMA).  The initial radio and mm-band observations
were presented in \citet{Soderberg+SN2011dh-I}.  Further radio
flux-density measurements as well as modeling of the light curve are
presented in a companion paper to the present one,
\citet{Krauss+SN2011dh-II}.

Although a yellow supergiant star was identified on pre-explosion {\em
  Hubble Space Telescope} ({\em HST}) images as a possible progenitor
\citep{vDyk+2011, Maund+2011},
subsequent work by \citet{Arcavi+2011b} suggested 
a smaller progenitor.
\citet{Soderberg+SN2011dh-I} examined the early data at radio through
X-ray wavelengths and concluded that these also suggest a small
progenitor star with a stellar radius of $\sim 10^{11}$~cm and were
not consistent with a radius of order $10^{13}$~cm as would be
expected for a yellow supergiant.
The yellow supergiant identified in the {\em HST} images would then
have to be either a binary companion of, or possibly a chance
superposition with, the true progenitor star.

\citet{ChevalierS2010} proposed that Type IIb supernovae (SNe~IIb) are
divided into two distinct subclasses.  The first, SN cIIb, have
compact progenitors, with stellar radii $\sim 10^{11}$~cm, (e.g.,
SN~2001ig, \citealt{Ryder+2004}; SN~2003bg, \citealt
{Soderberg+2006e}; and SN~2008ax, \citealt{Roming+2009}).  They are
characterized by high shock velocities of $\sim 0.1\,c$, and radio
light curves which show deviations from a power-law decay at late
times.  The second, SN eIIb, (e.g., SN~1993J, \citealt{SN93J-2};
SN~2001gd \citealt{Perez-Torres+2005}) have extended progenitors, with
stellar radii $\sim 10^{13}$~cm, and are characterized by somewhat
slower shock velocities and smooth radio light curves.

The size and expansion velocity of the shockfront is a basic
characteristic distinguishing different supernovae, and it is
therefore important to determine it observationally as directly as
possible.  In particular, to determine whether SN~2011dh conforms to
the characteristics of Chevalier \& Soderberg's
\nocite{ChevalierS2010} proposed Type SN~cIIb class requires measuring
the expansion velocity of the shock front.  Very
long-baseline-interferometry (VLBI) observations are the most direct
way of making this measurement \citep[see e.g.,][]{SN2009bb-VLBI,
  Brunthaler+2010a, SNVLBI-Bologna}.  Unlike the optical emission,
which mostly originates in the denser and more slowly moving inner
ejecta, the radio emission generally traces the fastest ejecta. The
radio emission is thought to originate in the region between the
forward and reverse shocks. In the particularly well-studied case of
SN~1993J, \citet{SN93J-4} show that there is a close relationship
between the outer boundary of the radio emission and the location of
the forward shock.

However, even with the high resolution afforded by VLBI, it is
generally difficult to measure the radius of the radio-emitting
region, and thus of the forward shock, especially early on in the
evolution of a supernova when its size is still small.  Chevalier
(1998; see also \citealt{ChevalierF2006})\nocite{Chevalier1998} has
shown that, for a supernova spectrum dominated by synchrotron
self-absorption (SSA), the radius of the emitting region, $R$,
(assumed spherical), as well as the magnetic field can be determined
from two observable quantities: the frequency and spectral luminosity
of the peak in the radio spectral energy distribution (SED)\@.  The
observables require only measurements of the total flux density at
different frequencies, and thus are not dependent on spatially
resolving the supernova.  In the case that significant free-free
absorption is present in addition to SSA, a lower limit on the
shockfront radius is obtained.

This calculation of the radius is fairly robust.  However, in addition
to the mentioned observables and the distance, $D$, which is usually
fairly well constrained, the calculation does also rely on
assumed values for two poorly known parameters.  The first of these,
$f$, is the filling factor of the radio emission, while the
second\footnote{Note that \citet{Krauss+SN2011dh-II},
  \citet{Soderberg+SN2011dh-I} and \citet{Chevalier1998} use the
  symbol $\alpha$ for this parameter, which symbol we use here to
  denote the radio spectral index.}, $\psi$, is the ratio of
relativistic electron energy density to magnetic energy density in the
post-shock region.

In the case of SN~2011dh, \citet{Soderberg+SN2011dh-I} showed that the
cm and mm-wave radio flux densities are consistent with a synchrotron
self-absorbed spectrum at $t \simeq 5$ and 17~days, and determined
that $R \simeq 3.7\times10^{15}$~cm at the latter epoch.  In our
companion paper, \citet{Krauss+SN2011dh-II} continue this work through
to $t = 92$~days, and present detailed EVLA monitoring of the
multi-frequency radio light curves, as well as modeling the light
curves to show that the evolution of $R$, as calculated from the radio
SED, is consistent with a power-law evolution with $R \propto
t^{0.9}$.

We report in this paper on VLBI observations of SN~2011dh obtained at
$t = 83$ and 179~days after the explosion, which have allowed us to
directly measure (or at least constrain) the angular size of SN~2011dh
at those epochs.  Early results from the first set of observations
have already been reported in \citet{SN2011dh_ATel}.

\section{Radio Light Curves}
\label{slightcurve}

\begin{figure*}
\centering
\epsscale{0.90}
\plotone{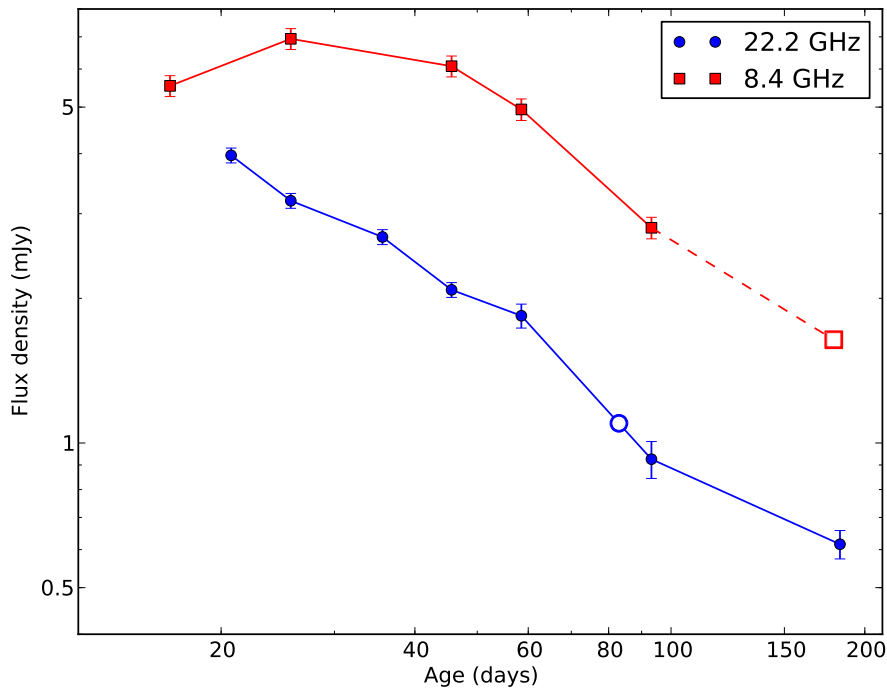}
\caption{Radio light curves of SN~2011dh at 22.2 and 8.4 GHz, as
  obtained from EVLA observations.  Flux density measurements are
  shown by solid circles and squares.  See text for details of how the
  flux density values were obtained.  With the exception of the one at
  age = 183~days, they were taken or interpolated from
  \citet{Krauss+SN2011dh-II}.  The open square represents the
  interpolated value at our first VLBI observing date and frequency
  (22.2~GHz at age = 83~days) while the open circle and dashed line show
  the extrapolation to our second VLBI session (8.4~GHz at age =
  179~days). The age of the supernova is calculated from the explosion
  date of 2011 May 31.6 UT.}
\label{flightcurve}
\end{figure*}

To set the context for, and aid in the interpretation of, the VLBI
results, we present in Figure~\ref{flightcurve} the radio light curves
of SN~2011dh at 22.2 and 8.4~GHz, the two frequencies at which we
carried out the VLBI observations.  The flux density values at
22.2~GHz were logarithmically interpolated in frequency between 20.5
and 25.0~GHz from those presented in \citet{Krauss+SN2011dh-II}, with
the exception of the last one at $t = 183$~days.  For that last one, we
obtained additional EVLA observations on 2011 November 30 at 20.5 and
25~GHz. The data reduction procedure was as described in
\citet{Krauss+SN2011dh-II}.  We obtained flux densities of of $722 \pm
72 \; \mu$Jy at 20.5~GHz and $551 \pm 55 \; \mu$Jy at 25~GHz, with the
uncertainties being intended as standard errors and dominated by an
estimated 10\% systematic contribution.
The value for the flux density at the time of the 22.2-GHz VLBI
observations at $t = 92$~days was then interpolated logarithmically first
in frequency and then in time between the adjacent values.

The 8.4-GHz measurements were also taken from
\citet{Krauss+SN2011dh-II}.  Near the time of the second VLBI
observations, the VLA was in the compact D array configuration, with
resolution insufficient to overcome confusion at 8.4 GHz, so a direct
measurement could not be obtained.  We therefore estimated the 8.4-GHz
flux density by first interpolating the 22.2-GHz values
logarithmically in time to $t = 179$~days, and then extrapolating to
8.4~GHz by assuming the same radio spectral index of $\alpha = -1$
(where $S_\nu \propto \nu^\alpha$) as was found appropriate for the
optically thin part of the spectrum through to $t = 92$~days
\citep{Krauss+SN2011dh-II}.  Although the optically-thin spectral
index has been seen to vary with time in some supernovae
\citep[e.g.,][]{SN93J-2, SN79C-shell}, large variations are not
expected, so there is no reason to think the model of
\citet{Krauss+SN2011dh-II} can not be extrapolated to $t = 179$~days, so
our estimate of the 8.4-GHz flux density at that time is of sufficient
accuracy for comparing to the flux density recovered from the VLBI
observations.

\section{VLBI Observations}
\label{sobs}

Our two sessions of VLBI imaging observations of SN~2011dh were carried
out using the high-sensitivity array, which consisted of the NRAO Very
Long Baseline Array (VLBA; $10 \times 25$-m diameter, distributed
across the United States), the NRAO Robert C. Byrd ($\sim$105~m
diameter) telescope and the Effelsberg (100~m diameter) telescope.
Both sessions were 13~hr in length, with the first using an observing
frequency of 22.2 GHz on 2011 August 21 UTC, and the second using
an observing frequency of 8.4 GHz on 2011 November 26 UTC.
The age of the supernova was 83 and 179~days, respectively, at the
midpoints of our two sessions.

In each set of observations, we included several ``geodetic blocks,''
where we observed 12-15 bright sources from the International
Celestial Reference Frame (ICRF) list of sources \citep{FeyGJ2009} over
a period of $\sim$45~minutes to measure the tropospheric zenith delay and
clock offsets at each antenna
\citep[see][]{BrunthalerRF2005,ReidB2004}. In these geodetic blocks we
recorded 8 intermediate frequencies (IFs) with frequency centers
spread over a $\sim$400~MHz range, with each IF covering a bandwidth
of 8~MHz.

For the observations of SN~2011dh we recorded a contiguous bandwidth
of 64~MHz in each of the two senses of circular polarization with
two-bit sampling\footnote{We note that parallel recordings were made
  using the higher bandwidth of 256 MHz per polarization,
  corresponding to a bit rate of 2 Gbit~s$^{-1}$, with the
  experimental new Mark 5C recording units.  Various technical
  difficulties with these recordings, however, rendered the correlated
  data unreliable, and we base the results in this paper only on the
  data recorded at 512~Mbit~s$^{-1}$ using the well-tried Mark 5B
  recording systems.}, for a total bit rate of 512~Mbit~s$^{-1}$.  We
used J1332+4722 (\objectname{ICRF J133245.2+472222}), 0\fdg{5} away
from SN~2011dh,
as a primary phase calibrator.  This source is an ICRF source
\citep{FeyGJ2009} with a position known to $\sim$70~\muas.  Any
positions in this paper are calculated by taking the position of
J1332+4722 to be RA = \RA{13}{32}{45}{24642}, decl.\ =
\dec{47}{22}{22}{6670} for J1332+4722.
For the first session, at 22~GHz, we used a cycle time of
$\sim$110~s, with $\sim$60~s duration on SN~2011dh and $\sim$50~s on
J1332+4722, while for the second session at 8.4~GHz, we used a
somewhat longer cycle time of $\sim$170~s, with $\sim$120~s spent on
SN~2011dh.  

We considered as a possible phase-reference source the weak nuclear
radio source of M51 \citep{Maddox+2007}, which is only 
$\sim$3\arcmin\ from SN~2011dh.  This nuclear source, however,
has a low 5-GHz peak brightness of $<$1~mJy, as well as having a steep
spectral index, and is therefore too weak to use for phase referencing
at our observing frequencies of 8.4 and 22.2 GHz.

During each of the observing runs, we also spent three $\sim$20-m
periods observing an astrometric ``check'' source, the quasar
\objectname{JVAS J1324+4743}, which was about 1\fdg{5}\ away from
our primary phase-reference source J1332+4722.  The purpose of the
observations of J1324+4743 was to check the quality of the
phase-referencing and also to provide a second astrometric reference
source.  This check source was observed using a similar
phase-referencing pattern as we used for SN~2011dh.

The VLBI data were correlated with the DiFX correlator
\citep{Deller+2011}, and the analysis carried out with NRAO's
Astronomical Image Processing System (AIPS) and ParselToungue
\citep{Kettenis+2006}. We calibrated both sets of observations using
standard procedures, making a correction for the dispersive
ionospheric delay using the AIPS task TECOR, and solving for the
zenith tropospheric delay on the basis of our geodetic-block
observations. We discarded any SN~2011dh visibility data obtained when
either of two the antennas involved was observing at elevations below
10\arcdeg.

The initial flux density calibration was done through measurements of
the system temperature at each telescope, and then improved through
self-calibration of the primary reference source J1332+4722.  This
source is slightly resolved, as can be seen on the images in the VLBA
calibrator list
data-base\footnote{http://www.vlba.nrao.edu/astro/calib}, where a
weak extension or second component is visible $\sim$2~mas to the
west-southwest of the peak.  We see a similar structure in the images
made from our data at both 22 and 8.4~GHz.  Our final amplitude and
phase calibration at both frequencies was derived using a CLEAN model
of this source, with the peak-brightness point in the image being placed at
the nominal coordinates given above\footnote{The deviation of the
  source geometry of J1332+4722 from a point source is small enough so
  that the effect of using a point model in the solutions for delay
  and delay rate made using FRING solutions is negligible.}.
Finally this calibration was interpolated to the intervening scans
of SN~2011dh.

\section{Results}
\subsection{Results from the 2011 August 24 Observations at 22 GHz}
\label{sresults-aug}

Unfortunately, the weather was generally poor for our first set of
observations, at 22 GHz, resulting in higher than usual noise levels:
the rms background brightness in a naturally-weighted image was
92~\muJb.  Nonetheless SN~2011dh was clearly detected, with a peak
brightness of 630~\muJb, which was 6.8 times the rms background.  As
mentioned above, early results from this epoch were reported
in \citet{SN2011dh_ATel}.

For marginally resolved sources, such as SN~2011dh, the best values
for the source size and VLBI flux density come from fitting models
directly to the visibility data, rather than from imaging.  We chose
as a model the projection of an optically-thin spherical shell of
uniform volume emissivity, with an outer radius of $1.25\times$ the
inner radius\footnote{Our results do not depend significantly on the
  assumed ratio between inner and outer radii, as the effect of
  reasonable variations in this ratio is considerably less than our
  stated uncertainties.  For a discussion of uncertainties on the
  shell-size obtained through \uv~plane modelfitting compared with
  those obtained in the image plane for the case of SN~1993J, showing
  that superior results are obtained using the former, see
  \citet{SN93J_Manchester}.}.
Such a model has been found to be appropriate for other radio
supernovae \citep[see e.g.,][]{SN93J-3, SN79C-shell}.  The Fourier
transform of this shell model is then fitted to the visibility
measurements by least squares.  For a partially resolved source such
as SN~2011dh, the exact model geometry is not critical, and our shell
model will give a reasonable estimate of the size of any circularly
symmetric source, with a scaling factor of order unity dependent on
the exact morphology \citep[see discussion in][]{SN93J-2}.

For the observations of SN~2011dh at $t = 83$~days, our best-fit
spherical shell model had a total flux density of $650 \pm
140$~$\mu$Jy (statistical and systematic uncertainty combined).
We note that the total flux density as interpolated from the VLA
measurements was 1100~$\mu$Jy (see Section~\ref{slightcurve} and
Figure~\ref{flightcurve}), with an estimated uncertainty of 10\%, and
is therefore higher than that recovered from the VLBI observations by
a combined $2.5\sigma$.  This discrepancy suggests that there is
likely some loss of phase-coherence in the VLBI observations, which is
not unexpected given the relatively poor weather.

We obtained a value of $0.11^{-0.11}_{+0.09}$~mas for the outer
angular radius of SN~2011dh at this epoch, where the listed
uncertainty is intended as a standard error, and consists of both the
statistical and systematic components added in quadrature.  We
estimated the systematic component as follows: the largest systematic
error most likely arises from the uncertainty in the antenna amplitude
gains, which are only imprecisely known but are correlated with the
fitted source size for marginally resolved sources, as well as from
the likely coherence losses mentioned above.  Although the effect of
the latter is difficult to estimate reliably, the first order effect
would be a reduction in the visibility amplitude.  We accordingly
estimated the effects of both amplitude-gain errors and coherence loss
from a small Monte-Carlo simulation where we randomly varied the
amplitude gains of the individual telescopes by 20\% rms, to arrive at
the uncertainty in the angular radius listed above.  The supernova is
therefore not significantly resolved at this epoch, and the $3\sigma$
upper limit on the outer radius was 0.38~mas.

We also take the fitted center position of the model as our best
estimate of the center-position of SN~2011dh.  For this epoch, this
position was RA = \RA{13}{30}{5}{105548} and decl.\ =
\dec{47}{10}{10}{92273}.
The statistical uncertainty on this position is $\sim$50~\muas,
however, a realistic position uncertainty will have substantial
contributions from unmodeled effects of the troposphere, ionosphere,
station position errors, and possible evolution of the reference
sources. We estimate the total uncertainty in the position relative to
that of J1332+4722 to be $\sim$70~\muas.  Our measured position
therefore agrees to within less than the combined uncertainties with
that measured by \citet{Marti-Vidal+2011d} at $t = 14$~days.

\subsection{Results from the 2011 November 26 Observations at 8.4 GHz}
\label{sresults-nov}

\begin{figure*}
\centering
\epsscale{0.75}
\plotone{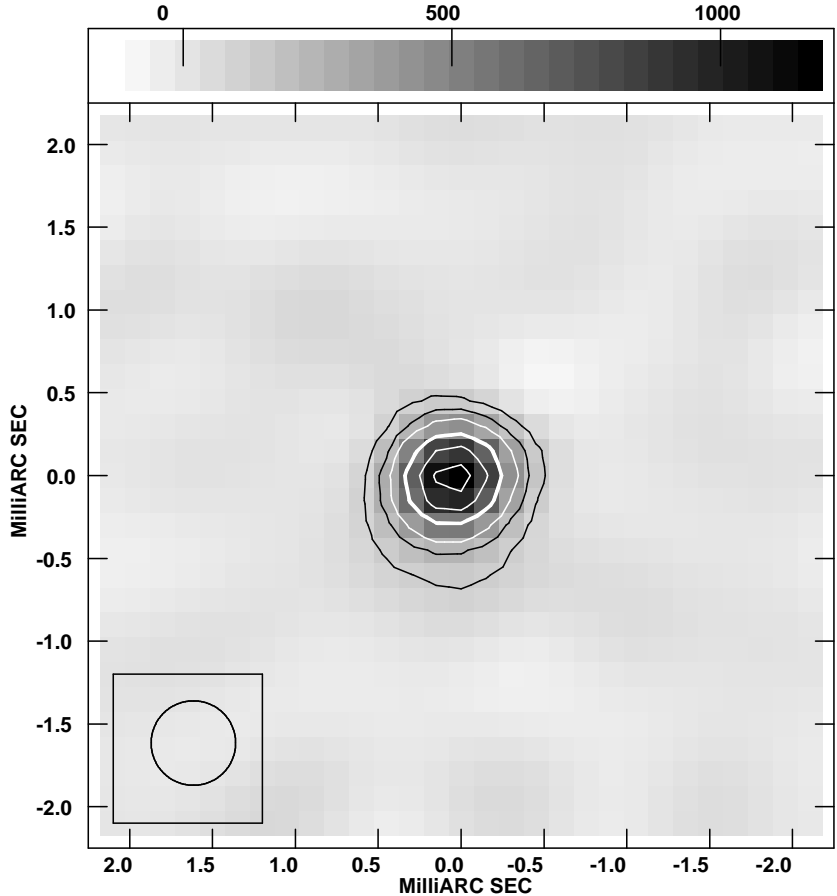}
\caption{The CLEAN image of SN~2011dh made from VLBI observations at
  8.4~GHz on 2011 November 26 UTC.  To obtain more robust imaging, we
  started with the square root of the visibility weights and then used
  uniform weighting.  The fitted CLEAN beam FWHM was $0.72 \times
  0.51$~mas at p.a.\ $-30$\arcdeg, and we super-resolved slightly by
  using a round CLEAN beam with FWHM of 0.51~mas, indicated at lower
  left.  The contours are drawn at 10, 20, 30, {\bf 50}, 70 and 90\%
  of the peak brightness, with the 50\% contour being emphasized.  The
  peak brightness was 1160~$\mu$\Jb, and the background rms brightness
  was 22~\muJb.  The greyscale is labeled in $\mu$\Jb.  North is up
  and east is to the left, and the coordinate origin is at the fitted
  center position of SN~2011dh, which was at RA
  \RA{13}{30}{05}{105548}, decl.\ \dec{47}{10}{10}{92273}.}
\label{fimage}
\end{figure*}

In the second set of VLBI observations at 8.4 GHz, SN~2011dh was
detected with good signal-to-noise ratio.  We show the VLBI image of
SN~2011dh in Figure~\ref{fimage}\@.  The fitted FWHM size of the CLEAN
beam was $0.72 \times 0.51$~mas at p.a.\ $-30$\arcdeg.  For
presentation we slightly super-resolved the image by restoring with a
round clean beam of FWHM 0.51~mas.  We emphasize that our measurement
of the angular size of SN~2011dh, as presented below, is entirely
independent of this super-resolution.  The peak brightness in this
image was 1160~$\mu$\Jb, while the rms background brightness was
22~\muJb.

Since SN~2011dh is only partly resolved, we again turn to fitting
models directly to the visibility measurements in order to accurately
measure the size of the source. Due to the good signal-to-noise ratio
obtained at this epoch, the source size could be fairly accurately
constrained by this method.  In the model-fitting, we used the
square-root of the uncertainties of the visibility measurements, which
procedure, though statistically less efficient, increases the
robustness of results in the likely case that the errors in the
visibility measurements are not purely statistical.  We again used a
spherical shell model with an outer radius 1.25 times the inner
radius.

From the model-fitting, we obtained a total flux density of $1540 \pm
190 \; \mu$Jy (an almost identical value was obtained from imaging).
This value is only 7\% below that of 1.64~mJy extrapolated for this
epoch and frequency from the EVLA measurements (see
Section~\ref{slightcurve} and Figure~\ref{flightcurve}), although we
note that the extrapolation EVLA value is somewhat uncertain in that
it depends on the errors in the individual flux density measurements
as well as the assumption that the radio spectral index does not vary
in time.  We find therefore that the flux density recovered from the
VLBI measurements is in reasonable agreement with that extrapolated
from the EVLA measurements, and that there is no reason to suspect
significant correlation losses.

We again also obtained a center position of SN~2011dh from this fit,
which was consistent within the uncertainties with that obtained for
our first epoch as well as with the position obtained at $t = 14$~days
by \citet{Marti-Vidal+2011d}.  The formal difference in position over
the 165-day interval between the Mart\'i-Vidal et al.\ observation and
the present one was $110 \pm 125$~\muas.

Finally, we obtained an outer radius for SN~2011dh of $0.25 \pm
0.08$~mas, with the uncertainties again including both statistical and
systematic contributions and derived as in the previous section.  At a
distance of 8.4~Mpc, this angular radius corresponds to an average
expansion velocity since the explosion of $21000 \pm 7000$~\kms.

To determine the quality of the phase referencing, upon which the
above determinations of the position and angular radius depend, we
examined the observations of our check source, J1324+4743.  From the
strictly phase-referenced data for the check source we found a total
flux density of 101~mJy, while for the phase self-calibrated data we
found a 36\% higher value of 137~mJy.  We fitted an elliptical
Gaussian model, appropriate for a marginally resolved QSO, directly to
the visibilities in a similar fashion as the shell model was fit to
the SN~2011dh data.  When fitting the strictly phase-referenced data,
we found that the fitted FWHM major axis was 25\% larger than when we
fit to the phase self-calibrated data.  This implies that, at least in
the case of our check source J1324+4743, the coherence losses due to
phase-referencing can result in an error of 25\% on the source size.
We note however, that J1324+4743 was almost three times farther from
the phase-reference source (J1332+4722) than was SN~2011dh.
Furthermore, the \uv~coverage for J1324+4743 was much poorer since we
only observed it for three brief periods.  We expect therefore that
the additional errors in the size of SN~2011dh due to coherence loss
are considerably smaller than 25\%, and thus considerably smaller than
the uncertainties from other causes discussed above.  We do not,
therefore, find any reason to suspect that coherence losses due to
poor phase-referencing would substantially increase the uncertainty we
give above for the angular size of SN~2011dh.

\section{Discussion}
\label{sdiscuss}

\begin{figure*}[ht]
\centering
\epsscale{0.85}
\plotone{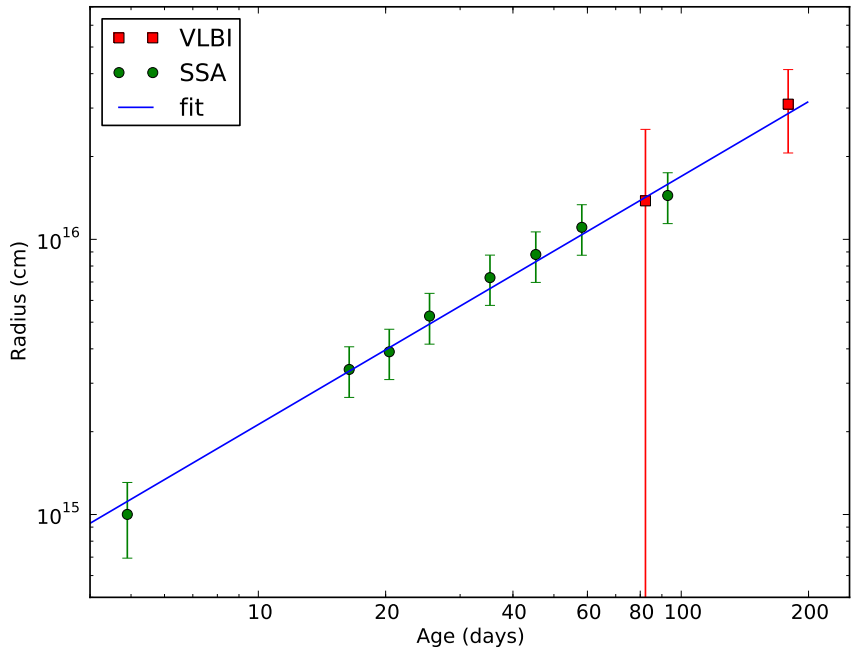}
\caption{The shock front radii of SN~2011dh as estimated by two
  independent methods.  In red, we plot the values derived from
  fitting spherical shell models to the VLBI visibility-data from this
  paper.  The plotted $1\sigma$ error bars include statistical and
  systematic contributions (see text Section~\ref{sobs} for details).
  In green, we plot the values calculated from the radio spectral
  energy distribution under the assumption that it is dominated by SSA
  (synchrotron self-absorption), with the radius values taken from
  \citet{Soderberg+SN2011dh-I} and \citet{Krauss+SN2011dh-II}, again
  with uncertainties including statistical and systematic components.
  The blue line represents our power-law fit to all the values, with
  $R = 5.85\times10^{15}\;(t/{\rm 30\, days})^{0.92}$~cm (see text).}
\label{frvst}
\end{figure*}

We obtained VLBI observations of the nearby Type IIb supernova 2011dh,
with the primary goal of obtaining a direct observational constraint
on the expansion speed by measuring the angular size of SN~2011dh.
VLBI observations are crucial, as they are the only means to directly
measure the size, expansion speed and perhaps the geometry of the
radio emission region.  We obtained an upper limit on the radius of
SN~2011dh at $t = 83$~days and a measurement of the radius at $t =
179$~days.  We plot these two values as a function of time in
Figure~\ref{frvst} (red squares).

As mentioned in the introduction, the radius of a supernova can also
be calculated from broadband, but not spatially-resolved, observations
of the radio SED (spectral energy distribution), under the assumption
of a spectrum dominated by SSA (synchrotron self-absorption). In our
companion paper Krauss et al.\ (2012; see also
\citealt{Soderberg+SN2011dh-I}) \nocite{Krauss+SN2011dh-II} we
used such observations of the SED to calculate the shockfront radius
for SN~2011dh through to $t = 92$.  In particular, the radius is
calculated following Chevalier \& Fransson (2006; see also
\citealt{Chevalier1998})\nocite{ChevalierF2006}:

\begin{equation}
\label{eqrssa}
\begin{split}
  R =& \; 4.0\times10^{14} \; \psi^{-1/19} \;
  \left({{f}\over{0.5}}\right)^{-1/19}
  \left({{S_{\nu\pk}} \over {\rm mJy}}\right)^{9/19} \\
  & \times \left({{D \over {\rm Mpc}}}\right) ^{18/19} 
  \left({{\nu\pk}\over{\rm 5 \; GHz}}\right) ^{-1} \rm cm,
\end{split}
\end{equation}

\noindent where $S_{\nu\pk}$ is the observed flux density at the peak
of the synchrotron spectrum, which occurs at frequency $\nu\pk$, $D$
is the distance, and $f$ and $\psi$ are, as mentioned above, the
filling factor and the ratio of the energy density in relativistic
particles to that of the magnetic field, respectively.

For $f$ we have some observational constraints at least in the
well-studied case of another SN IIb, SN~1993J, where the
well-resolved VLBI images suggest a very spherical shell structure
with a shell thickness of 20\% to 33\% of the outer radius
\citep[e.g.,][]{SN93J-2, SN93J_Manchester, Marti-Vidal+2010a}.  Such a
geometry would have a gross value of $f$ in the range of 49\% to 70\%,
although it is possible that small scale clumpiness within this
overall geometry could reduce $f$ below those values.  For $\psi$, the
commonly used value is $\psi = 1$, corresponding to equipartition.
Here, and in our companion paper \citep{Krauss+SN2011dh-II} we
therefore take $f = 0.5$ and $\psi = 1$.

The radius of SN~2011dh's forward shock was calculated from the SED as
obtained from EVLA and SMA observations using this method at various
times between $t = 4$ and 92~days by \citet{Soderberg+SN2011dh-I}
and \citet{Krauss+SN2011dh-II}.  We plot these values also in
Figure~\ref{frvst} (green circles).

Assuming that the supernova ejecta and circumstellar material have
power-law density profiles, then evolution of the forward shock radius
with time follows a power-law form and can be expressed simply as $R
\propto t^m$, where $m$ is known as the deceleration parameter.  If
the circumstellar density is, as expected for a stellar wind, $\propto
R^{-2}$, and the outer portions of the ejecta are also characterized
by a power-law density distribution with index $n$, then it can be
shown that $m = (n - 3)/(n - 2)$ \citep{Chevalier1982a}.  Indeed, we
found in \citet{Krauss+SN2011dh-II}, that at least up to $t = 92$~days
the evolution of SN~2011dh's radius was consistent with $m = 0.9$,
implying that $n \simeq 12$.

In order to determine $m$ from our various measurements of SN~2011dh's
radius, we performed a weighted least-squares fit to the complete set
of values for $R$, including both those determined from our VLBI
measurements and those from fitting the SED. The fitted line is also
plotted in Figure~\ref{frvst} (in blue).  As can be seen in
Figure~\ref{frvst}, the expansion of SN~2011dh is in fact well
described by a power-law.  The fit gives $m = 0.92 \pm 0.03$ and a
radius at $t = 30$~days of $(5.85 \pm 0.13) \times 10^{15}$~cm, where
the uncertainties are statistical ones derived from the scatter of the
fit.

In addition to the power-law nature of the expansion, it can be seen
from Figure~\ref{frvst} that there is excellent agreement between the
radii measured with VLBI with those calculated from the radio SED
under the assumption of SSA.  This agreement provides strong
validation for the latter method of calculating the shockfront
radius.  Indeed, SN~2011dh represents so far the best example for
directly comparing the radii of the shock wave determined in these two
different fashions.

The radii derived from fits to the SED are uncertain due to a number
of systematic factors.  Firstly, they are calculated from the fitting
of a model SSA spectrum to the observed flux density values.  In the
case of SN~2011dh, although the SSA model is an excellent fit over a
wide spectral range, the fit is not exact \citep[see e.g.,][]
{Krauss+SN2011dh-II}, resulting in estimated systematic uncertainties
of 15\% and 5\%, respectively, on the values of $\nu\pk$ and
$S_{\nu\pk}$, which result in a $\sim$15\% systematic uncertainty on
the calculated radius.  In Figure~\ref{frvst}, the plotted error bars,
intended as standard errors, include this systematic component on the
values of $R$ calculated from the SED.  We performed a Monte-Carlo
simulation where we varied all the radius measurements according to
their standard errors, and found an rms scatter of the derived values
of $m$ of 0.10\@. 
It can be seen, however, that any such systematic errors do not seem
time-dependent, in other words likely have only a small effect on the
derived value of $m$.  We therefore conservatively estimate the
systematic $1\sigma$ uncertainty on our fitted value of $m$ at 0.10.

The radii calculated from the SED also depend, as noted above, on the
poorly known parameter $\psi$, for which we took the equipartition
value of unity.  \citet{Soderberg+SN2011dh-I}, however, found that for
SN~2011dh the X-ray measurements, in conjunction with those in the
radio, suggested deviations from equipartition, with $\psi \sim 30$.
Can we constrain $\psi$ by comparing the values of the $R$ as
calculated from the SED with those from VLBI which are independent of
$\psi$?

As can be seen from Equation \ref{eqrssa} above, the former values of
$R$ in fact depend only weakly on $\psi$.  In particular, the change
of $\psi$ from our assumed value of unity to $\psi = 30$, as suggested
by \citet{Soderberg+SN2011dh-I}, would decrease the calculated values
of $R$ by 16\%, producing only a small change in Figure~\ref{frvst}.
Using such slightly lower values of the $R$ determined from the SED
(but with the original values of $R$ from VLBI) does not change the
fitted value of $m$ by more than the statistical uncertainty, and
results in a slightly, but not significantly, poorer fit of the
power-law expansion.  We therefore conclude that the measurements are
not yet of sufficient accuracy to usefully constrain $\psi$.

As we have shown, the deceleration parameter, $m$, for SN~2011dh seems
robustly determined at $m \simeq 0.9$, which implies only slightly
decelerated expansion, and consequently a relatively steep radial
density profile in the outer ejecta, with power-law index $n \simeq
12$. We note that \citet{Chevalier1982a} found that for self-similar
models with $m = 0.9$ and $n = 12$ the ratio of outer shock to inner
shock radii was 1.24, almost identical to our adopted value of 1.25.

The deceleration of SN~2011dh may show a pattern similar to
that seen for SN~1993J, where $m \simeq 0.9$ was found for
approximately the first year \citep{SN93J-2, Marti-Vidal+2010a},
albeit with somewhat smaller velocities than SN~2011dh.
We note that the expansion velocities SN~2011dh's shock front, as
determined from the radio emission, are approximately twice as large
as the highest velocity of $\sim$17000~\kms\ observed in the optical
spectrum \citep[from the blue edge of the H$\alpha$ emission
  line;][]{SilvermanFC2011, Arcavi+2011b}.
This is not unexpected as the high-velocity shocked H is likely to
be non-radiative \citep{ChevalierS2010}, and the ejecta should be
characterized by a steep density profile \citep{BergerKC2002,
  ChevalierF2006}.

Our observations also constrain the proper motion of the geometrical
center of SN~2011dh: between the EVN observations of
\citet{Marti-Vidal+2011d} at $t = 14$~days and our observations at $t =
179$~days the observed proper motion corresponds to a velocity of $9600
\pm 10500$~\kms, not significantly different from zero.  We note that
large proper motions of the center of the emission region are not
expected: the best-determined case is SN~1993J, where a peculiar
motion of only $320 \pm 160$~\kms\ was observed \citep{SN93J-1}.

As mentioned, \citet{ChevalierS2010} have suggested that type IIb
supernovae are divided into two categories according to whether the
progenitor was extended or compact.  Since SN~2011dh had a compact
progenitor \citep{Soderberg+SN2011dh-I, Krauss+SN2011dh-II}, it would
belong to the latter category.  According to \cite{ChevalierS2010}
this category is characterized by a high expansion velocity as well as
by variations in the light curve, and presumably also the deceleration,
due to episodic mass-loss from the progenitor.  SN~2011dh's relatively
high expansion velocity has already been noted, and is confirmed by
our VLBI observations.  It will, however, be important to continue
observations of SN~2011dh's radio SED as well as VLBI imaging for as
long as possible to determine whether the remainder of the predictions
are borne out, and to increase our sample of supernovae with
observationally determined expansion curves as well as radio and X-ray
light curves, since supernovae as nearby and as radio-bright as
SN~2011dh are rare events.

\acknowledgements 
\noindent{Research at York University was partly supported by NSERC.
A. B. was supported by a Marie Curie Outgoing International Fellowship
(FP7) of the European Union (project number 275596).  L.C. is a Jansky
Fellow at the NRAO.  We thank NRAO for scheduling these
target-of-opportunity observations.  Our results are partly based on
observations with the 100-m telescope of the MPIfR
(Max-Planck-Institut f\"ur Radioastronomie) at Effelsberg.  We made
use of the Swinburne University of Technology software correlator,
developed as part of the Australian Major National Research Facilities
Programme and operated under license.  In addition, we also made use
of NASA's Astrophysics Data System Bibliographic Services.}

\bibliographystyle{hapj}
\bibliography{mybib1,sn2011dh-temp}

\end{document}